\documentclass[aps,pre,twocolumn,longbibliography]{revtex4-1}
\usepackage{amssymb,amsmath}
\usepackage{graphicx}
\usepackage{algorithm}
\usepackage{algpseudocode}
\usepackage[%
    breaklinks=true,
    colorlinks=true,
    linkcolor=blue,
    urlcolor=blue,
    citecolor=blue
]{hyperref}

\begin{document}

\title{Joint effect of advection, diffusion, and capillary attraction
on the spatial structure of particle depositions from evaporating
droplets}

\author{K.S.~Kolegov$^{1,2,3}$}

\email{konstantin.kolegov@asu.edu.ru}

\author{L.Yu.~Barash$^{3,4}$}

\affiliation{$^1$ Astrakhan State University, 414056 Astrakhan, Russia}

\affiliation{$^2$ Volga State University of Water Transport, Caspian Institute of Maritime and River Transport, 414014 Astrakhan, Russia}

\affiliation{$^3$ Landau Institute for Theoretical Physics, 142432 Chernogolovka, Russia}

\affiliation{$^4$ National Research University Higher School of Economics, 101000 Moscow, Russia}

\begin{abstract}
A simplified model is developed, which allows us to perform computer
simulations of the particles transport in an evaporating droplet
with a contact line pinned to a hydrophilic substrate. The model
accounts for advection in the droplet, diffusion and particle
attraction by capillary forces. On the basis of the simulations, we
analyze the physical mechanisms of forming of individual chains of
particles inside the annular sediment. The parameters chosen
correspond to the experiments of Park and Moon
[\href{https://doi.org/10.1021/la053450j}{Langmuir {\bf 22}, 3506
(2006)}], where an annular deposition and snakelike chains of
colloid particles have been identified. The annular sediment is
formed by advection and diffusion transport. We find that the close
packing of the particles in the sediment is possible if the
evaporation time exceeds the characteristic time of diffusion-based
ordering. We show that the chains are formed by the end of the
evaporation process due to capillary attraction of particles in the
region bounded by a fixing radius, where the local droplet height is
comparable to the particle size. At the beginning of the
evaporation, the annular deposition is shown to expand faster than
the fixing radius moves. However, by the end of the process, the
fixing radius rapidly outreaches the expanding inner front of the
ring. The snakelike chains are formed at this final stage when the
fixing radius moves toward the symmetry axis.
\end{abstract}


\maketitle

\section{Introduction}

The processes of heat and mass transfer occurring in droplets and films are of interest
in various applications such as fuel droplet evaporation
and combustion in engines~\cite{Sazhin2018_6498,Gambaryan-Roisman2019_115},
interaction of droplets with surfaces of varying wettabilities in ink-jet printing~\cite{Shikhmurzaev2012_082001},
obtaining stable ultrathin film surfaces on the basis of polar liquids~\cite{Gordeeva2017_531},
removing nanoparticles from a solid surface~\cite{MAHDI201513}, and many other applications.

\begin{figure}[h!]
 \includegraphics[width=0.95\linewidth]{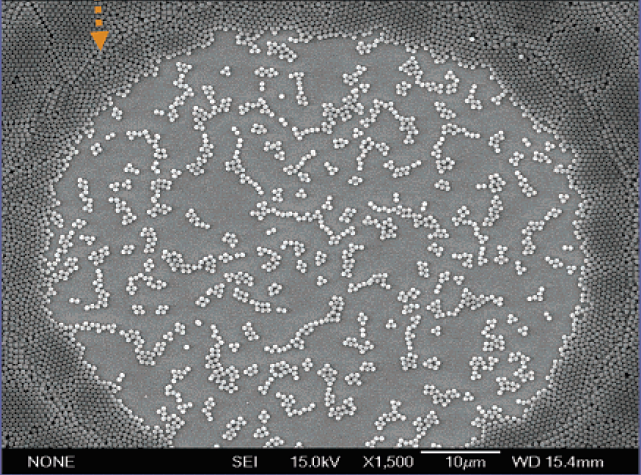}%
 \caption{\label{fig:ParkMoonExperiment}
 Experimental observation of individual chains of particles within the annular
 deposition (reprinted with permission from~\cite{Park20063506}, copyright 2006, American Chemical
 Society).}
 \end{figure}

One of the most important and actively discussed problems is
connected to studying structures of colloidal particles, which
emerge on the surface of an evaporating sessile droplet and remain
on the substrate after
drying~\cite{Larson20141538,Thiele2014,Erbil2015,Sefiane2014,Brugarolas2013,Brutin2018,Tarafdar2018,Pauchard201832}.
One of the examples is the effect of evaporative contact line
deposition, the so-called coffee-ring
effect~\cite{Deegan1997,Deegan2000,Mampallil2018}. While a droplet
is drying on the substrate, capillary flows carry the colloid
particles toward the three-phase boundary. In this case, formation
of an annular deposition is observed if the contact line was pinned
throughout the entire process. According to the experimental
results, coffee-rings may have an inner structure~\cite{Marin2011},
and colloid particles can also merge into chains and other geometric
shapes~\cite{Park20063506} (Fig.~\ref{fig:ParkMoonExperiment}).
However, the existing models which describe this
phenomenon~\cite{Deegan2000, Fischer2002, TarasevichModernPhys2011,
Ozawa2005} do not allow predicting the inner structure and location
of the particles relatively to each other.

Effects of this kind are studied experimentally
in~\cite{Park20063506,Marin2011,Zhang2013,
Yunker2011,Yunker2013,Mayarani2017_18798,Li2008_14266,Callegari2018,Govor2004_061609,Govor2004_4774,Govor2009_488}.
Mar\'{\i}n \textit{et al}. primarily explained the mechanism of
these effects by convective transfer and diffusion~\cite{Marin2011}.
Park and Moon obtained circular depositions and stains as a result
of evaporating picoliter droplets based on water ink and different
solvent mixtures~\cite{Park20063506}. Such depositions are
characterized by a well-ordered hexagonal structure of particles. In
case of the annular deposition, the images show the presence of
independent chains of particles within the ring.

The role of the Marangoni flow in case of a heated substrate is
discussed in~\cite{Zhang2013}. Ternary mixtures of solvent, polymer
and semiconductor nanocrystal were investigated
in~\cite{Miller20141665}. Yunker \textit{et al}. studied the
influence of particle shape on the resulting structure of
deposition~\cite{Yunker2011,Yunker2013}. Depositions of microgel
soft particles are characterized by a gradual order-to-disorder
transition~\cite{Mayarani2017_18798}, which is not abrupt like the
hard colloid depositions~\cite{Mayarani2017_18798,Marin2011}. The
transition from order to disorder is also observed in the structures
obtained by a dip-coating
technique~\cite{Kaplan2015,Mehraeen2015,Colosqui2013}. Li \textit{et
al}. used solvents with dendrimers and obtained annular structures
of a periodic thickness~\cite{Li2008_14266}. Callegari \textit{et
al}. studied the influence of active matter (moving bacteria) on the
dynamics of the coffee-rings growth~\cite{Callegari2018}. A series
of papers is dedicated to the formation of nanoparticle chains along
the contact line of micron-size
droplets~\cite{Govor2004_061609,Govor2004_4774,Govor2009_488}.

The network deposition patterns inside the annular sediment have been obtained experimentally
for a drying sessile droplet in~\cite{LiLanWang2017}, where
a tailored substrate with a circular hydrophilic domain is used.
The evolution of a dry patch is found to be divided into three stages:
rupture initiation, dry patch expansion, and drying of the residual liquid.
The authors argue that in their case the capillary attraction is important,
while the Marangoni flows are considerably large
only in the first phase of evaporation, when the contact angle is relatively large.

The capillary
attraction of the particles in liquid films was studied
in~\cite{KralchevskyNagayama1994,Denkov19923183,Denkov199326,Park20117676,Lotito2017217,TangCheng2018}.
Based on such interaction between the particles, a lithographic
method of self-assembly of nanostructures was
developed~\cite{Liddle20043409}.

There are several discrete models, which explicitly describe the
dynamics of each particular particle. These models are usually based
on Monte Carlo or molecular dynamics
methods~\cite{Liao2000_84,gundabala2003,LebedevStepanov2013132,Mehraeen2015,Shi2016,Akkus2018,TangCheng2018}.
Some Monte Carlo based models are lattice
models~\cite{Zhang2016650,Kim2011,jung2014}, and some are
not~\cite{Petsi2010,Callegari2018,Kolegov2019012043}. The flow
velocity is often calculated analytically for simple particular
cases~\cite{LebedevStepanov2013132,Kim2011,Petsi2010,jung2014,crivoi2014,Kolegov2019012043}.
The obtained velocity field is taken into account for calculations
of the particles dynamics. Callegari \textit{et
al}.~\cite{Callegari2018} use a constant velocity value. Mobile
bacteria are modeled by the Monte Carlo method~\cite{Callegari2018}.
Lebedev-Stepanov and Vlasov took account of the adhesion of
particles and roughness of the
substrate~\cite{LebedevStepanov2013132}. The formation of branched
nanoparticle aggregates in the drying sessile droplet is modeled
in~\cite{Zhang2016650}, where the two-dimensional kinetic Monte
Carlo Ising-like approach is used. The results pertain to those
experiments, where nanoparticles are considered, and the coffee ring
effect is absent. The branched aggregates begin to form when the
thickness of the ultrathin film reaches several angstroms and local
ruptures appear. On such a scale, the process can be accompanied by
the formation of other structures, such as networks, worm-like
structures and so on~\cite{Stannard2011}. The
models~\cite{Kim2011,crivoi2014} describe the formation of annular
depositions in the process of droplet evaporation using a cubic
lattice. Petsi \textit{et al}. compared the Monte Carlo and lattice
Boltzmann equation methods for a semi-cylindrical geometry of the
droplet and also provided a numerical description of the process of
particles deposition on the substrate as a result of the transfer by
the compensatory fluid flow and their Brownian
motion~\cite{Petsi2010}. The computational experiment was carried
out for different wetting angles and three-phase boundary modes.
Jung \textit{et al}. used a model with a circular lattice to
describe the coffee-ring effect mathematically~\cite{jung2014}. A
computational method which allows one to calculate the capillary
attraction forces between particles in a thin film is presented
in~\cite{Zhou20063692}.

The objective of the present work is to analyze physical mechanisms
of the formation of standalone particle chains located within an
annular monolayer deposition~\cite{Park20063506}
(Fig.~\ref{fig:ParkMoonExperiment}), where the particle size is of
the order of micrometers. Our hypothesis is that this effect is
caused by the particles' capillary attraction, which occurs at
different times in those areas of the droplet where the thickness of
the liquid layer reduces to the size of particles. To verify this
mechanism, we employ a simplified mathematical model, which accounts
for joint consideration of advection, diffusion, and capillarity and
allows us to perform computer simulations of the transport of
particles.

\section{Methods}

\subsection{Problem formulation}

\begin{figure}[h!]
 \includegraphics[width=0.99\linewidth]{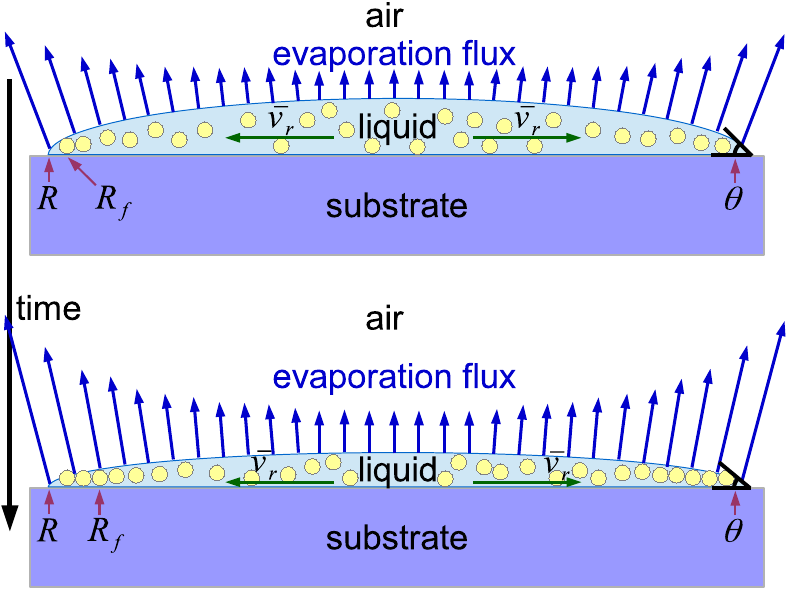}%
 \caption{\label{fig:problemDefinition} Sketch to the problem definition.}
 \end{figure}

Let us consider an evaporating sessile droplet of colloid solution
(Fig.~\ref{fig:problemDefinition}). The radius of the spherical
particles of $\mathrm{SiO_2}$ $r_p \approx$ 0.35~$\mu$m~\cite{Park20063506}.

We estimate the sedimentation time for a single particle
by dividing the droplet height $h_0=R\theta/2\approx 5$ $\mu$m
by the Stokes velocity $U_0 = 2r_p^2 \Delta\rho g/ 9\eta \approx 0.5$ $\mu$m/s~\cite{Russel1991},
where $\Delta\rho = 1.65$ g/cm$^3$ is the density difference
between the particles and the water, $g$ is acceleration due
to gravity, and $\eta$ is the fluid viscosity.
Hence, the sedimentation time is of the order of $t_s \approx 10$ s.
The advection further increases the sedimentation time,
which exceeds the evaporation time $t_{max} \le 10$ s.
For this reason, we do not take account of sedimentation.

Although the particles of silicon dioxide may acquire a charge during
their chemical synthesis~\cite{Behrens2001}, we assume that
the screening is significant due to a large concentration of ions
in the aqueous solution.
For this reason, we disregard the electrostatic interactions~\cite{Linse2005,Liang2007151}.

The shape of a sessile droplet profile is a spherical cap due to the
predominance of capillary forces over gravity because of the small
volume of liquid ($V\approx 72$ pl~\cite{Park20063506}). Such
geometry facilitates working with a cylindrical coordinate system
($r,\varphi, z$). The direction of axis $r$ is parallel to the
substrate. The symmetry center of the droplet corresponds to $r=0$.
Axis $z$ is perpendicular to the substrate. The surface of the solid
base corresponds to  $z=0$. The three-phase boundary is fixed
throughout the entire process at $r = R$. The contact line radius
is $R\approx$ 60~$\mu$m. The substrate is made from a
hydrophilic material, and the droplet is flattened. Its contact
angle $\theta \leq$ 10$^\circ$~\cite{Park20063506}. In such cases,
one or two layers of particles are formed~\cite{Park20063506}. In
this work, we only consider the case with a monolayer formation from
particles.

We take into account radial
flow of the liquid, directed from the symmetry axis toward the
droplet periphery. This compensatory flow is known as capillary flow
and is a result of intense evaporation near the contact line
(Fig.~\ref{fig:problemDefinition}).
We do not take account of circulating flows of the
liquid, because the liquid layer is thin (see Sec.~\ref{subSubSec:Marangoni}).

\subsection{Description of the model}
\label{subsec:Model}

\subsubsection{Capillary attraction}
\label{subSubSec:CapillaryAttraction}

It is convenient to use both polar ($r, \varphi$) and Cartesian
coordinates ($x, y$). Since thickness of the liquid layer $h$ in the
flattened droplet is much smaller than $R$, the particle dynamics
can be described in a horizontal plane, $xy$. Despite the fact that
we study the case with a fixed three-phase boundary, the $xy$ plane
also has a movable boundary, $R_f$. Let the fixing radius $R_f$
define a boundary, where the particle size and the local droplet
height are comparable ($d_p\approx h$), provided that $R_f < R$.
Here $d_p = 2r_p$. In
a thin droplet, the particles cannot reach the contact line, because
the local droplet height is very small in the vicinity of the
contact line. The particles will not move farther than $R_f$, since
beyond this boundary  $h < d_p$. The surface tension forces will
restrain them. Let us write the approximate expression for the shape
of the droplet surface~\cite{Popov2005}

\begin{equation}\label{eq:ParabolicDropShape}
h(r,t) = \theta (t) \frac{R^2 - r^2}{2R},
\end{equation}
where radial coordinate is $r = \sqrt{x^2 + y^2}$.
From~\eqref{eq:ParabolicDropShape} we derive the dependence of the fixing radius on time
\begin{equation}\label{eq:fixing_radius}
  R_f (t) = \sqrt{R^2 - \frac{4 r_p R}{\theta (t)}}.
\end{equation}

The droplet height and the contact angle reduce during the evaporation process,
and, hence, the fixing radius $R_f(t)$ is changed as well (Fig.~\ref{fig:problemDefinition}).
It should be noted
that $R_f$ is changed insignificantly for the majority of time. Only at the end of the process,
this radius starts shrinking. At $t \to t_\mathrm{max}$ the value of $R_f \to 0$. Contact angle $\theta$ decreases
linearly over time, $\theta \to 0$ at $t \to t_\mathrm{max}$~\cite{PhysRevE.83.051602},
where $t_\mathrm{max}$~is the time of full evaporation.
This dependence can be expressed as follows:
\begin{equation}\label{eq:ConatactAngleEvolution}
\theta(t) = \theta_0 \left( 1 - \frac{t}{t_\mathrm{max}} \right),
\end{equation}
where $\theta_0$ is the initial value of the contact angle~\cite{Popov2005}.
The exact value of $\theta_0$ is not determined in the experiment~\cite{Park20063506}.
It is only known that it does not exceed 10$^\circ$ for the case of the hydrophilic substrate.
For the calculations, we will use $\theta_0 =$ 10$^\circ$.

\begin{figure}[h!]
 \includegraphics[width=0.95\linewidth]{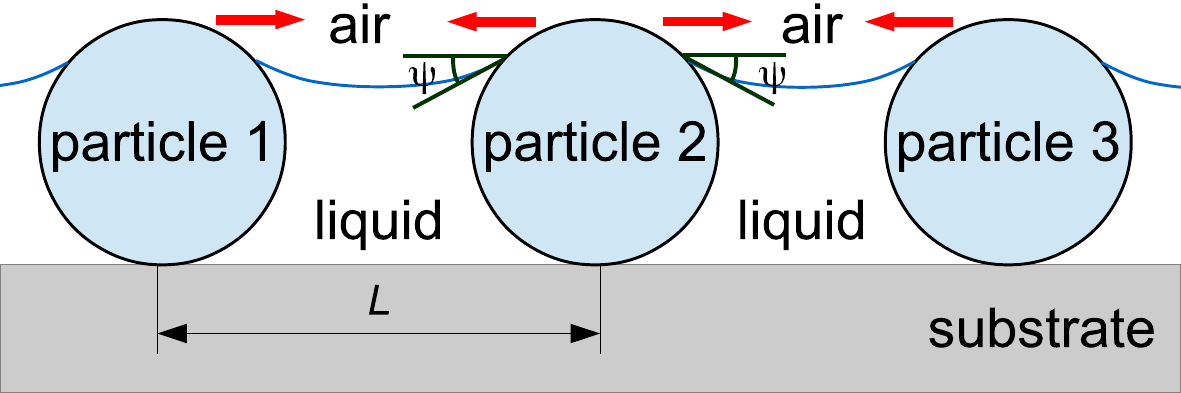}%
 \caption{\label{fig:capillaryAttraction}
 A schematic of particles capillary attraction.}
 \end{figure}

In the region of boundary $R_f$, the particles are subject to the capillary
attraction~\cite{Denkov19923183,KralchevskyNagayama1994,Lotito2017217}
\begin{equation}\label{eq:CapillaryAttractionForce}
  F \approx \frac{2\pi \sigma Q_1 Q_2}{L},
\end{equation}
where $L$~is the distance between particles, and $Q_k$ is the
capillary charge of the $k$th particle, and $\sigma$ is the surface
tension coefficient of the liquid.
Expression~\eqref{eq:CapillaryAttractionForce} resembles Coulomb's
law. Here, $Q_k$ characterizes the ability of a particle to deform
the free surface of the liquid, $Q_k = r_p\sin \psi_k$. The radius
of any particle is assumed to be constant value $r_p$. The meniscus
slope angle $\psi_k$ may vary depending on the location of a
particle. To carry out assessment, let us consider the average of
$\psi$ (Fig.~\ref{fig:capillaryAttraction}). In this case, $Q_1 =
Q_2 = Q$, where $Q = r_p \sin \psi$. Then,
expression~\eqref{eq:CapillaryAttractionForce} can be written as
follows:
\begin{equation}\label{eq:CapillaryAttractionForceSimplified}
  F \approx \frac{2\pi \sigma Q^2}{L}.
\end{equation}
Let us study a three-particle system as an example
(Fig.~\ref{fig:capillaryAttraction}) and assume that angle $\psi =$
1$^\circ$.

We assume that the left and the right particles are attached to the
substrate. Let us determine the displacement of the central particle
over a time step of $\delta t=$ 10$^{-4}$~s. The choice of the time
step value is discussed in Sec.~\ref{subsec:Algorithm}. Let the
distance between the first and second particles be $L_{12} = 11
r_p$, and the distance between the second and third particles be
$L_{23} = 12 r_p$. Now, let us determine the values of $F_{12}$ and
$F_{23}$ by substituting $L_{12}$ and $L_{23}$
in~\eqref{eq:CapillaryAttractionForceSimplified}. The second
particle displacement caused by capillary forces is
$$\delta l_{\mathrm {cap}}=\frac{F_{23}-F_{12}}{m_p} \frac{\delta t^2}{2}.$$
Mass of a particle $m_p = V_p \rho_p\approx$ 0.5~pg, where $V_p$ is
the volume of a spherical particle, and $\rho_p$ is the density of
silicon dioxide. We obtain the value of relation $\delta
l_{\mathrm{cap}}/L_{12}\approx -1$. This means that the second
particle shifts tightly to the first particle, since the distance
between them is smaller and the capillary effect is stronger. It
will occur more rapidly at a larger angle $\psi$.

\subsubsection{Advection}

Based on the mass conservation law~\cite{Deegan2000} and taking into
consideration the lubrication approximation $\sqrt{ 1 + (\partial h
/ \partial r)^2 } \approx 1$ we express the height-averaged velocity
of radial flow of the liquid $\bar v_r(r,t)$,
\begin{equation}\label{eq:velocity}
\bar v_r \approx - \frac{1}{rh} \int\limits_0^r \left(
\frac{J}{\rho_l} + \frac{\partial h}{\partial t} \right) r\,dr,
\end{equation}
where density of the liquid $\rho_l \approx 10^3$ kg/m$^3$, and
$J=J(r,t)$ is the local evaporation flux with a dimension of
kg/(m$^2$~s).

If we consider the evaporation as only a diffusion of vapor into air
and assume the process is quasistationary, then, by solving the
Laplace equation for the distribution of vapor concentration $\Delta
c = 0$, we find $J$. In this approximation, $J$ only depends on $r$.
The boundary condition at the liquid-vapor interface is $J(r) = D_v
\left.(\mathbf{n} \cdot \nabla c)\right|_{z=h}$, where $\mathbf{n}$
is the normal vector to the droplet surface, and $D_v$ is the
coefficient of vapor diffusion. Analytical solution
\begin{equation}\label{eq:evaporationFluxDensityAnalitical}
J(r) \approx \frac{2}{\pi} \frac{D_v (c_s - c_\infty)}{\sqrt{R^2 -
r^2}}
\end{equation}
had been obtained earlier for $\theta\to 0$~\cite{Deegan2000,Popov2005,Larson2005a}.
Here $c_s$ is the concentration of saturated vapor, and $c_\infty$
is the concentration of vapor in the surrounding air far from the droplet surface.

Let us substitute~\eqref{eq:ParabolicDropShape},
\eqref{eq:ConatactAngleEvolution} and
\eqref{eq:evaporationFluxDensityAnalitical} in~\eqref{eq:velocity}
and then integrate. Considering the fact that the mass of the
droplet $m$ is equal to the mass of the entire evaporated
liquid~\cite{Popov2005},
$$m = \frac{\pi \rho_l R^3
\theta_0}{4} = 2\pi \int\limits_0^R \int\limits_0^{t_\mathrm{max}}
J r\,dr\,dt = 4D_v (c_s - c_\infty) R t_\mathrm{max},$$ we find
 \begin{equation}\label{eq:velocityAnalytical}
 \bar v_r= \frac{R}{4\tilde r (t_\mathrm{max}-t)}\left[  \frac{1}{\sqrt{1-\tilde r^2}} - \left( 1-\tilde r^2 \right)
 \right],
\end{equation}
where $\tilde r = r/R$~\cite{Popov2005,Larson2005a}. Displacement of
the particle by the radial fluid flow over one time step is $\delta
l_{\mathrm{adv}} = \bar v_r \delta t$.

Equation~\eqref{eq:velocityAnalytical} is a rough approximation for
an actual fluid velocity, so the suggested model is only of a
phenomenological nature. The process can be considered as
quasistationary when $R^2 / D_v \ll
t_\mathrm{max}$~\cite{Popov2005}. In order to further refine the
suggested model, one has to consider $J$, which depends on time $t$,
as well. For example, dependencies $J(r,t)$ are suggested
in~\cite{MurisicKondic2008,Barash2009,Cazabat20102591,SemenovBrutinPRE2017}.

\subsubsection{Marangoni flows}
\label{subSubSec:Marangoni}

It can be shown that the capillary flow dominates the Marangoni effect in the fixing radius region.
As seen in Eq.(\ref{eq:velocityAnalytical}), the velocity of capillary flow
substantially increases both on approaching the contact line
and during the final phase of the evaporation process.
Since for large contact angles the fixing radius (\ref{eq:fixing_radius}) is close to the contact line,
the capillary flow dominates the Marangoni flow there.
The velocities of Marangoni flow are proportional to $\theta^2$,
and decrease rapidly during the final phase of the evaporation process.
Hence, at small contact angles, the capillary flow dominates the circulatory one as well.

It should be noted that the thermocapillary instability is known to
occur only when the Marangoni number $\mathrm{Ma}= -\sigma^\prime
\Delta T h_0 / (\eta \kappa)$ exceeds a certain critical value. Here
$\Delta T$ is the temperature difference between the apex and the
substrate, $\sigma^\prime = d\sigma/dT$, and $\kappa$ is the thermal
diffusivity. The threshold value for a flat fluid film is about
$80$~\cite{Pearson1958}. The onset of the Marangoni convection has
been also confirmed in~\cite{McDonaldWard2012} for the particular
case of droplets with contact angle $\theta = 90^\circ$. Using the
known estimates of the evaporation flux~\cite{HuLarson20021334}, we
obtain the value $\mathrm{Ma}\approx 0.4$ for $\theta = 10^\circ$
and $\mathrm{Ma}\approx 0.03$ for $\theta = 3^\circ$. Although the
onset is not known for droplets with small contact angles, it is
possible that for such small values of $\mathrm{Ma}$ the
thermocapillary flow does not occur at all.

For this reason, we do not take into account Marangoni flows in the
simulations.

\subsubsection{Diffusion}

Using the generalized Stokes-Einstein
equation~\cite{Yiantsios2006,Russel1991}, we estimate the diffusion
coefficient of the particles $D \approx 2\cdot 10^{-13}$ m$^2$/s.
The diffusion displacement distance is $\delta l_{\mathrm{dif}} =
\sqrt{2 D \delta t} \approx$ 6~nm. We obtain the relation: $\delta
l_{\mathrm{cap}}/ \delta l_{\mathrm{dif}} \approx 615$. Hence,
capillary attraction of the particles prevails over diffusion, which
can be neglected on achievement of fixing radius $R_f$ by the
particle. The characteristic time of diffusion-based ordering of
particles is $t_d = d_p^2 / D \approx$ 2.45 s. The evaporation time
of the droplet depends on the temperature, humidity, air pressure,
concentration of colloidal particles, and other parameters. The
precise value of evaporation time $t_\mathrm{max}$ is not given
in~\cite{Park20063506}. For a pure liquid, there is an approximation
 $$t_\mathrm{max} \approx
\frac{\rho_l h_0 R}{\pi D_v c_s \left(1-H\right)\left(0.27 \theta +
1.3\right)} \approx 0.3 \text{ s,}$$ where the coefficient of
diffusion is $D_v \approx 2.4 \cdot 10^{-5}$ m$^2$/s, concentration
of saturated vapor $c_s \approx 17.3\cdot 10^{-3}$
kg/m$^3$~\cite{Larson20141538}. The value of the contact wetting
angle is $\theta = \pi/ 18$ and height of the flattened droplet $h_0
\approx R \tan(0.5\theta) \approx$ 5 $\mu$m. In the
experiment~\cite{Park20063506}, relative humidity of the surrounding
air is $H\approx 0.4$. A number of
experiments~\cite{Okuzono2010,Kim2018} demonstrate that the
evaporation rate can decrease with increasing concentration of
colloidal particles. For this reason, we consider $t_\mathrm{max}$
as a parameter. In calculations, we will use $t_\mathrm{max}=$ 1~s
$< t_d$ and $t_\mathrm{max}=$ 10~s $> t_d$.

Brownian motion of the particles is simulated using a Monte Carlo
method. We denote a random angle between a particle displacement and
the $x$ axis as $\alpha$, $\alpha \in [-\pi; \pi)$. Then the
displacement of a particle is described by the vector $(\delta
x,\delta y)^T = (\delta l_{\mathrm{dif}} \cos \alpha,\delta
l_{\mathrm{dif}} \sin \alpha)^T$.

\subsection{The algorithm}
\label{subsec:Algorithm}

The pseudocode of the algorithm is given in the Appendix.
We use the Mersenne Twister generator
in order to generate pseudorandom numbers~\cite{Matsumoto19983}.
Efficient implementations for the current computer architectures are
contained in modern software libraries for random number
generation~\cite{rngsselib2011,rngavxlib2016,prand2014}.

\begin{figure}[h!]
 \includegraphics[width=0.95\linewidth]{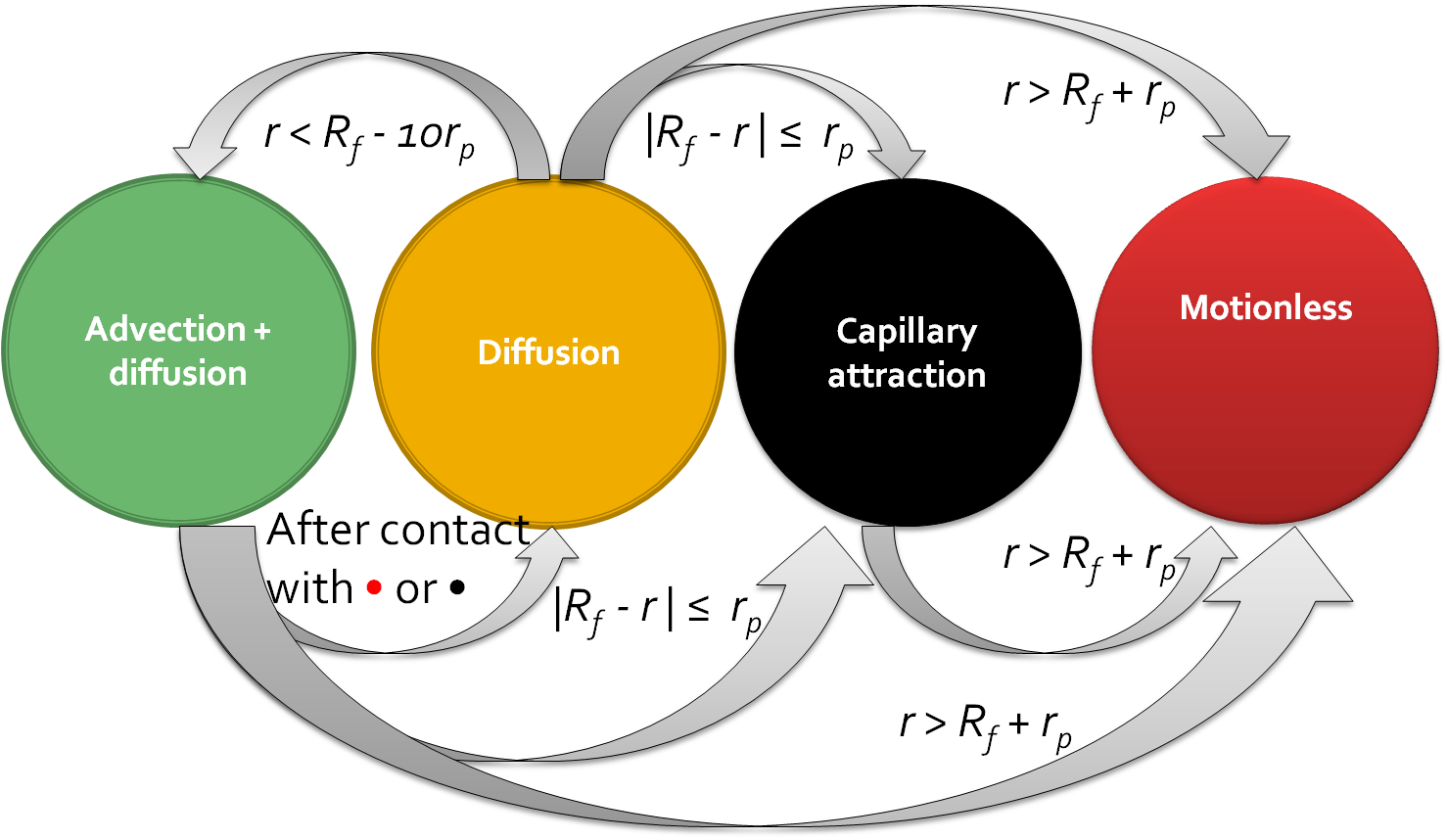}%
 \caption{\label{fig:FiniteStateMachine}
 Particle states and transition rules (color online).}
\end{figure}

The particles can be in different states during the calculation.
This approach resembles a finite state machine
(Fig.~\ref{fig:FiniteStateMachine}). The Supplemental Material video
[\href{http://link.aps.org/supplemental/10.1103/PhysRevE.100.033304}{77}]
shows results of the simulations and displays the states by
different colors. The particles subject to advection and diffusion
are marked in green. Yellow particles are only subject to diffusion.
Black particles are affected by capillarity. Red particles are the
ones that stopped their motion and are attached to the substrate.
Conventionally, below we use this color notation of the states,
because it is identical to the color notation in the Supplemental
Material video.

Initially, according to the color designation, the particles are
green by default. A particle becomes black on reaching fixing radius
$|R_f - r| \leq r_p$. If the particle offsets beyond boundary $R_f$
toward $R$ when $r > R_f + r_p$, it becomes red.

The particles within the forming ring are more affected by diffusion
than by fluid flow.
This can be seen, e.g., in the supplemental video in~\cite{Yunker2011}.
We assume that a green particle becomes yellow if its attempt
to move was failed due to a collision with a red or black particle.
In this context, collision is an overlap of the particles, i.e., in
the two-dimensional model representation, an overlap of the circles
of radius $r_p$. The particles are of a solid material and are not
deformed. The regions of space occupied by the particles cannot
cross even partially.

If the yellow particle has moved away by a small distance from $R_f$
due to the Brownian motion toward the symmetry axis (e.g., $r < R_f
- 5d_p$), it becomes green again. In other words, the liquid flow
starts moving it toward the periphery again.

The capillarity of particles on the basis of assessments using
formula~\eqref{eq:CapillaryAttractionForceSimplified}
in~Sec.~\ref{subSubSec:CapillaryAttraction} will be simulated as
follows. Over one time step, a black particle is attracted and moves
closely to the nearest black or red particle in the neighborhood
(within the surrounding region of a radius $R_n$). For a
calculation, we take $R_n = 20 r_p$~\cite{Park20117676}. The value
of time step $\delta t=$ 10$^{-4}$ s was chosen on the basis of a
series of computation experiments to satisfy the Einstein relation
for the mean square displacement $\langle \delta L_{\mathrm{dif}}^2
\rangle = 2D t_\mathrm{max}$, where $\delta L_{\mathrm{dif}}$ is the total displacement
during the period $t_\mathrm{max}$, and the averaging is performed
over all particles.

We do not consider collisions of the particles. Since the Monte
Carlo method updates the particle motion at each time step, its
stochastic character excludes a possibility to describe collisions
of particles given such a small time step $\delta t$. The diffusion
displacement $\delta l_\mathrm{dif}$ during a time step is about 1\%
of the particle size. Particle displacement by advection $\delta
l_\mathrm{adv}$ usually does not exceed 1.5\% of the particle size.
This results in a negligibly small portion of colliding particles
during one step.

\section{Results and discussion}

\begin{figure*}
 \includegraphics[width=0.85\linewidth]{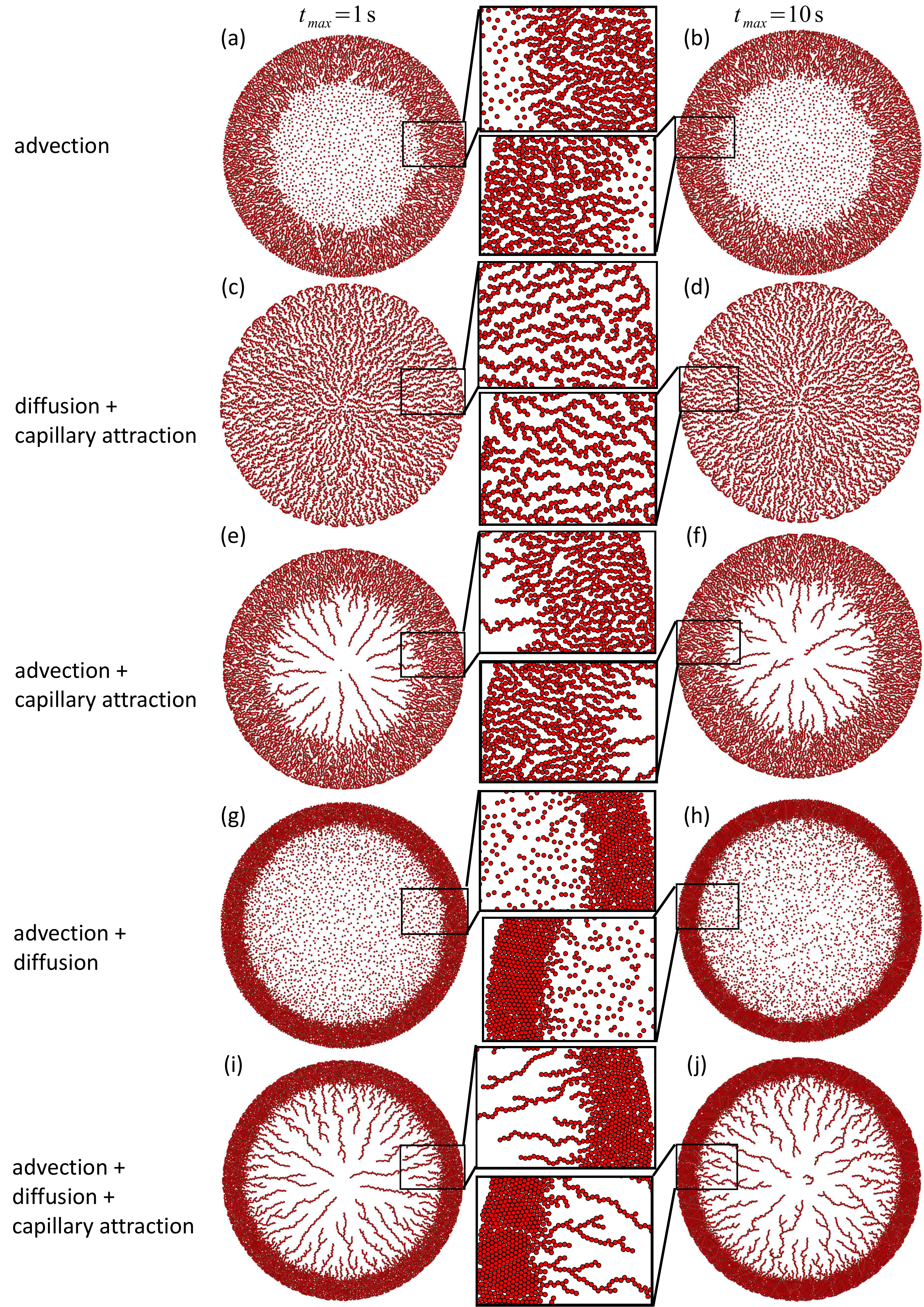}
 \caption{\label{fig:structure} Final structures obtained at different combinations of effects taken into account
and values of $t_\mathrm{max}$.}
 \end{figure*}

\begin{figure*}
     \includegraphics[height=4.25cm]{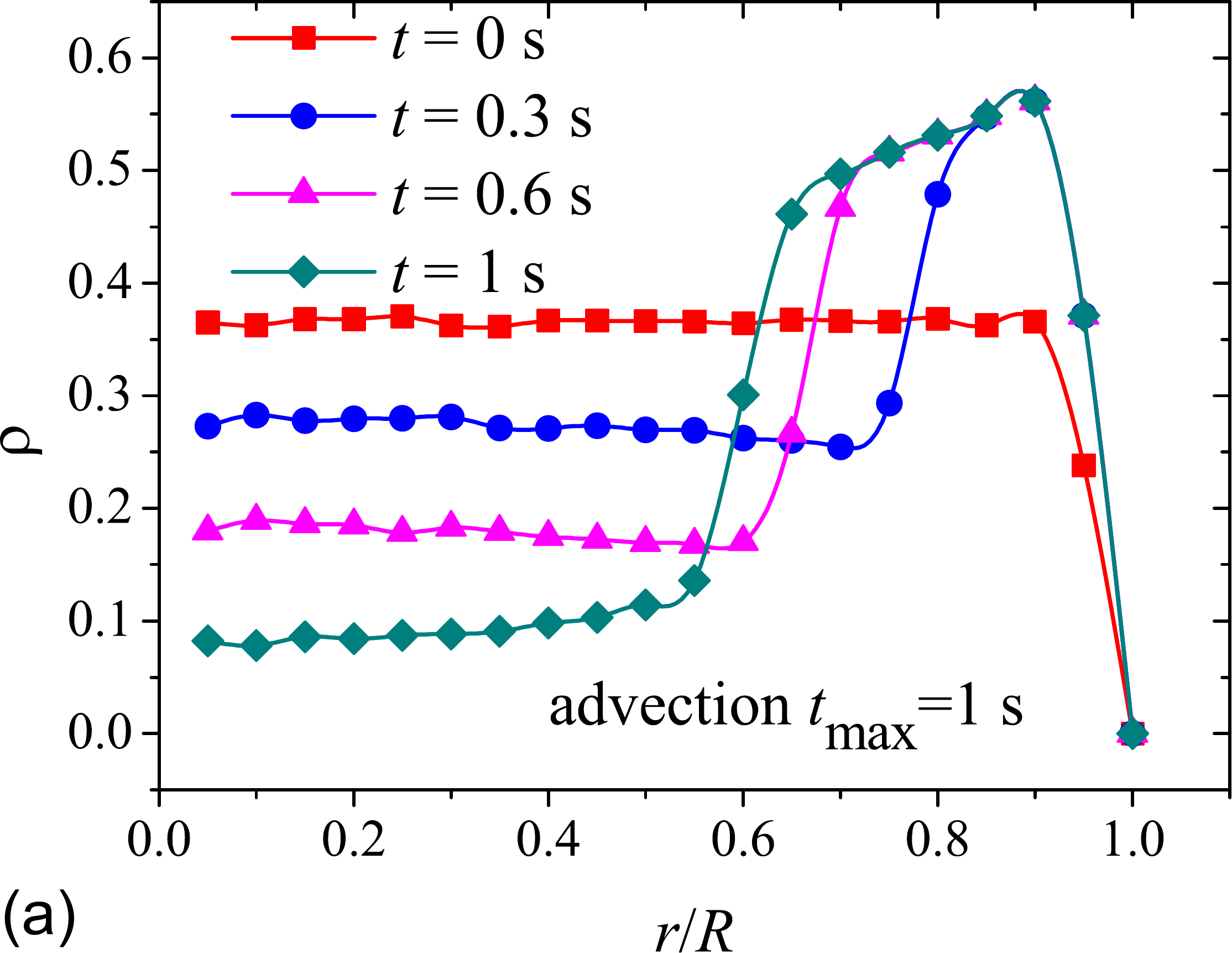}\hspace{30pt}
     \includegraphics[height=4.25cm]{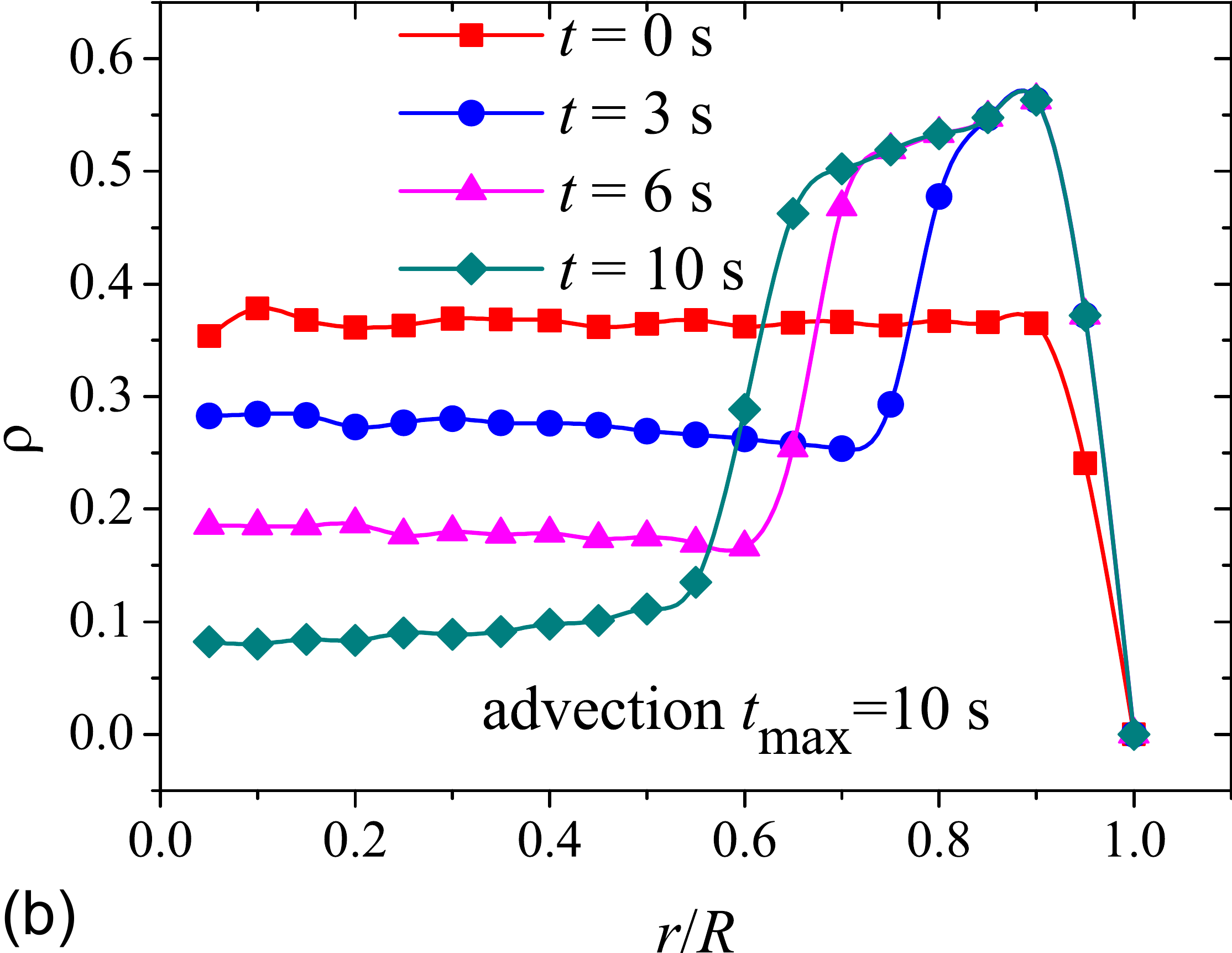}\\ \vspace{10pt}
     \includegraphics[height=4.25cm]{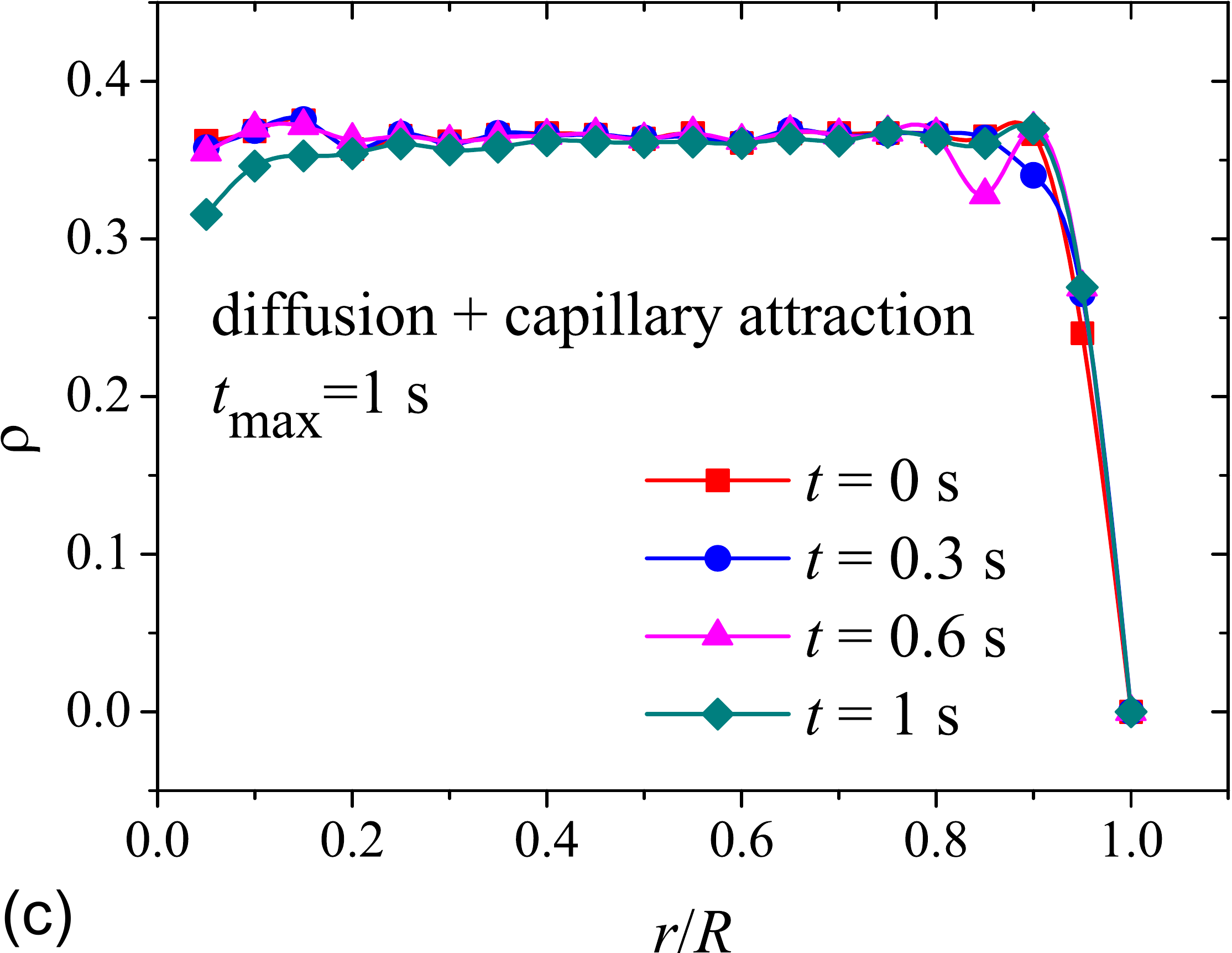}\hspace{30pt}
     \includegraphics[height=4.25cm]{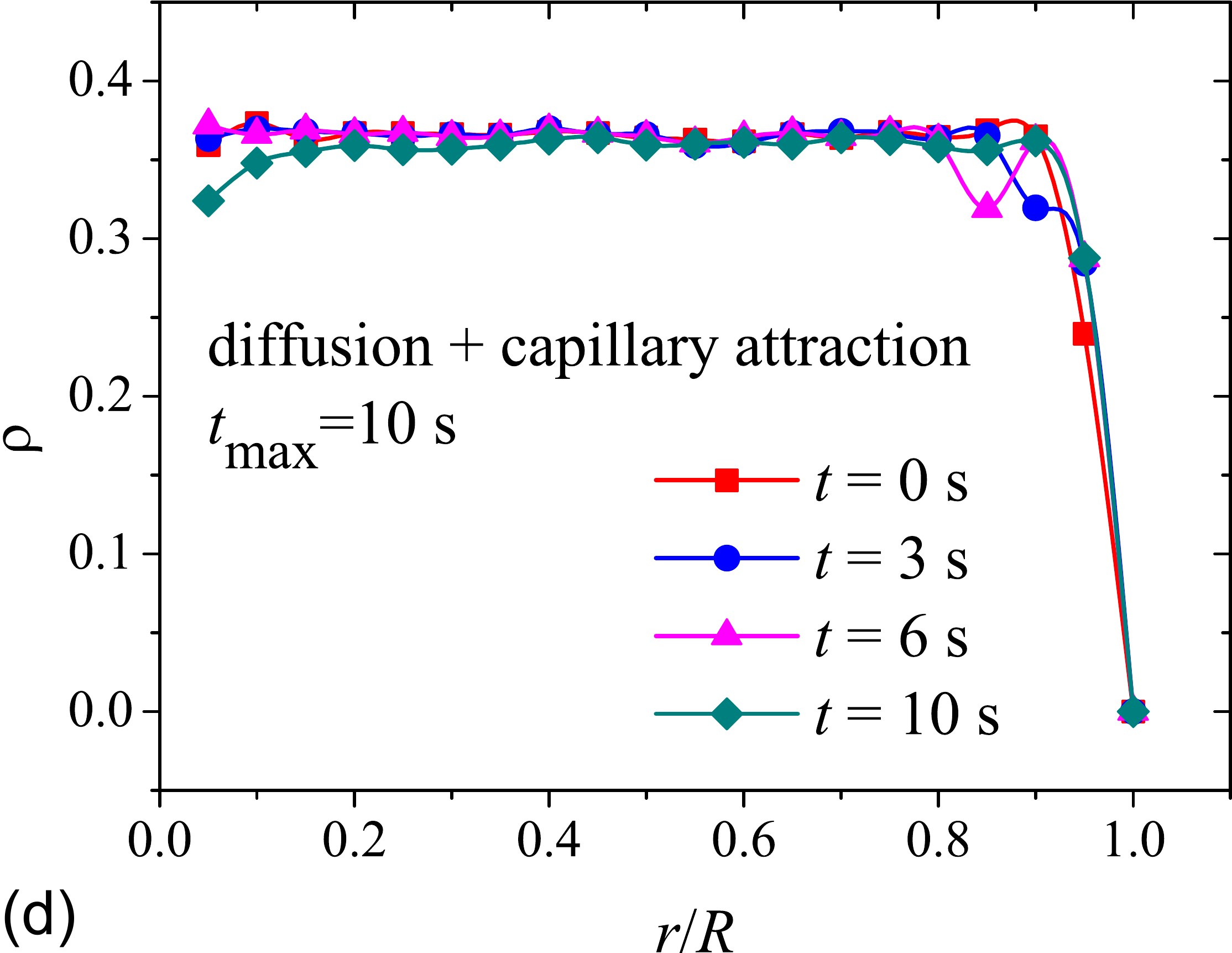}\\ \vspace{10pt}
     \includegraphics[height=4.25cm]{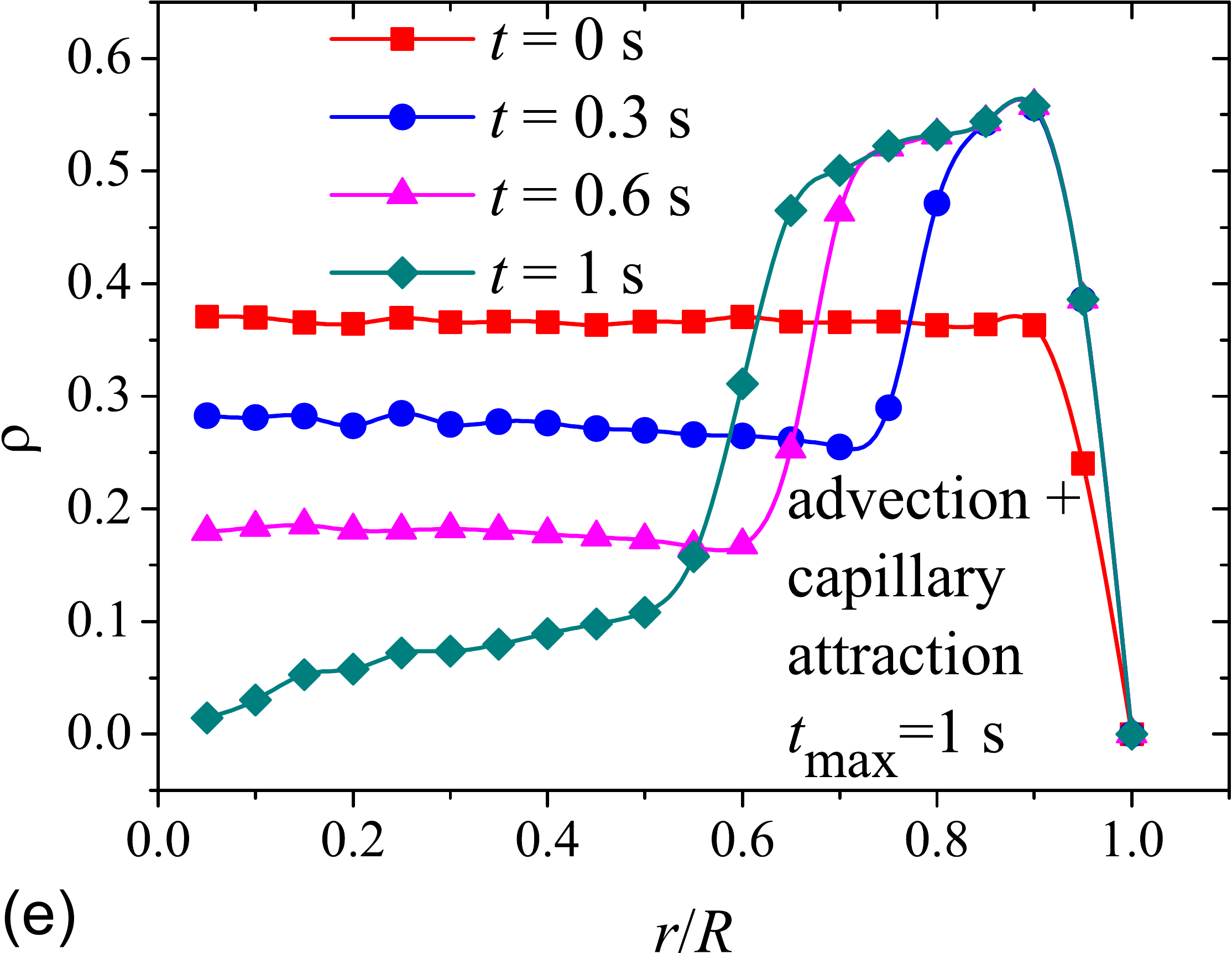}\hspace{30pt}
     \includegraphics[height=4.25cm]{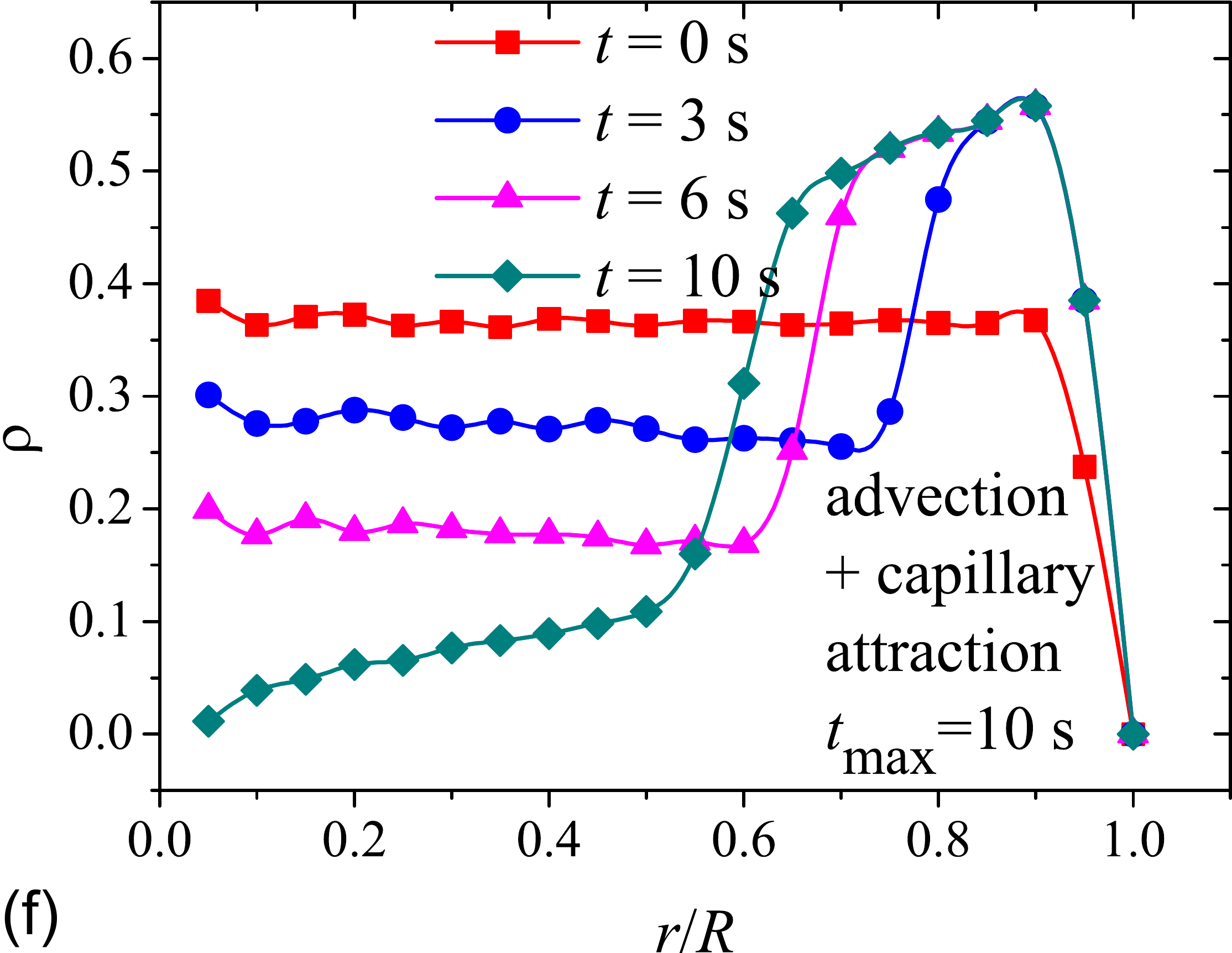}\\ \vspace{10pt}
     \includegraphics[height=4.25cm]{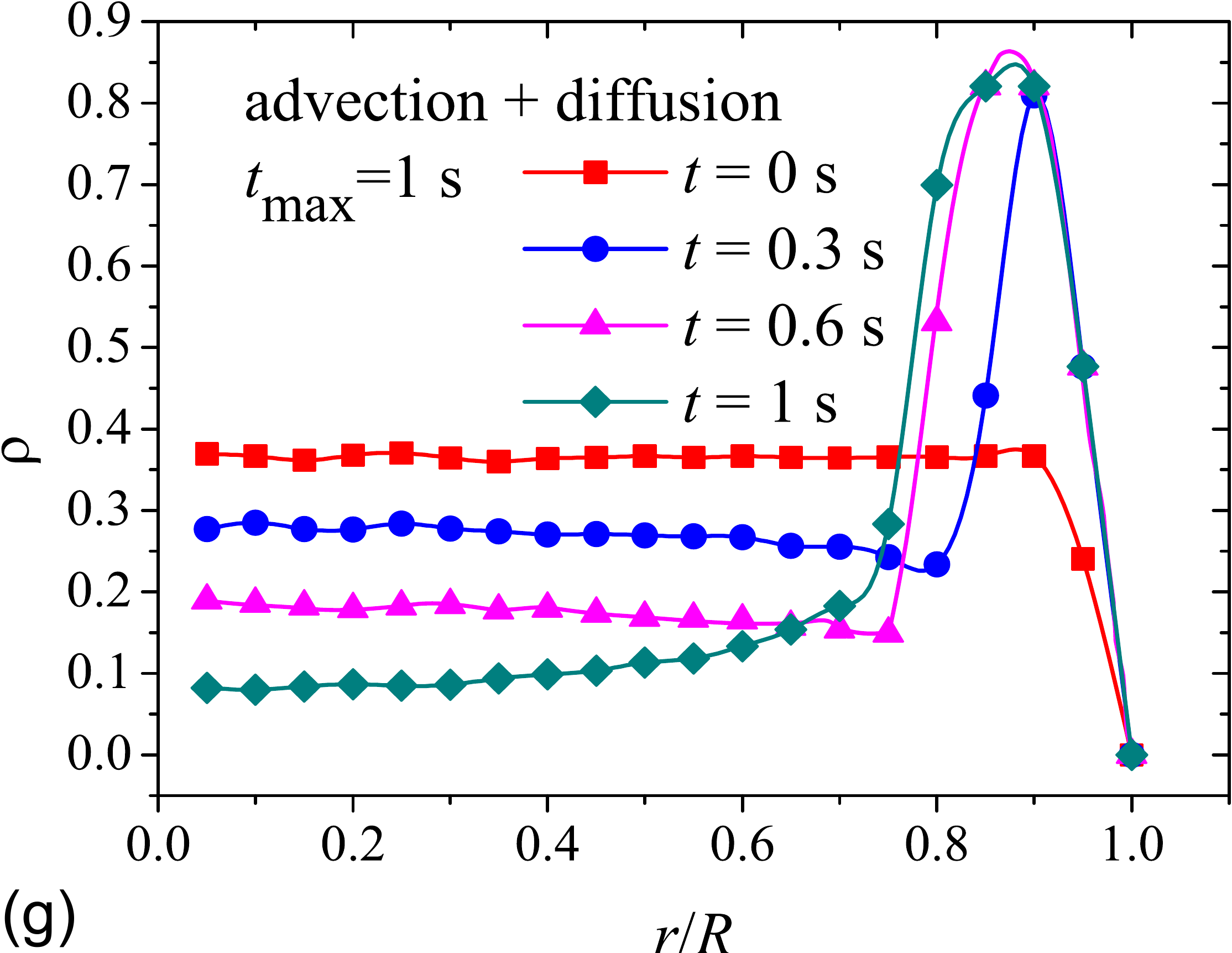}  \hspace{30pt}
     \includegraphics[height=4.25cm]{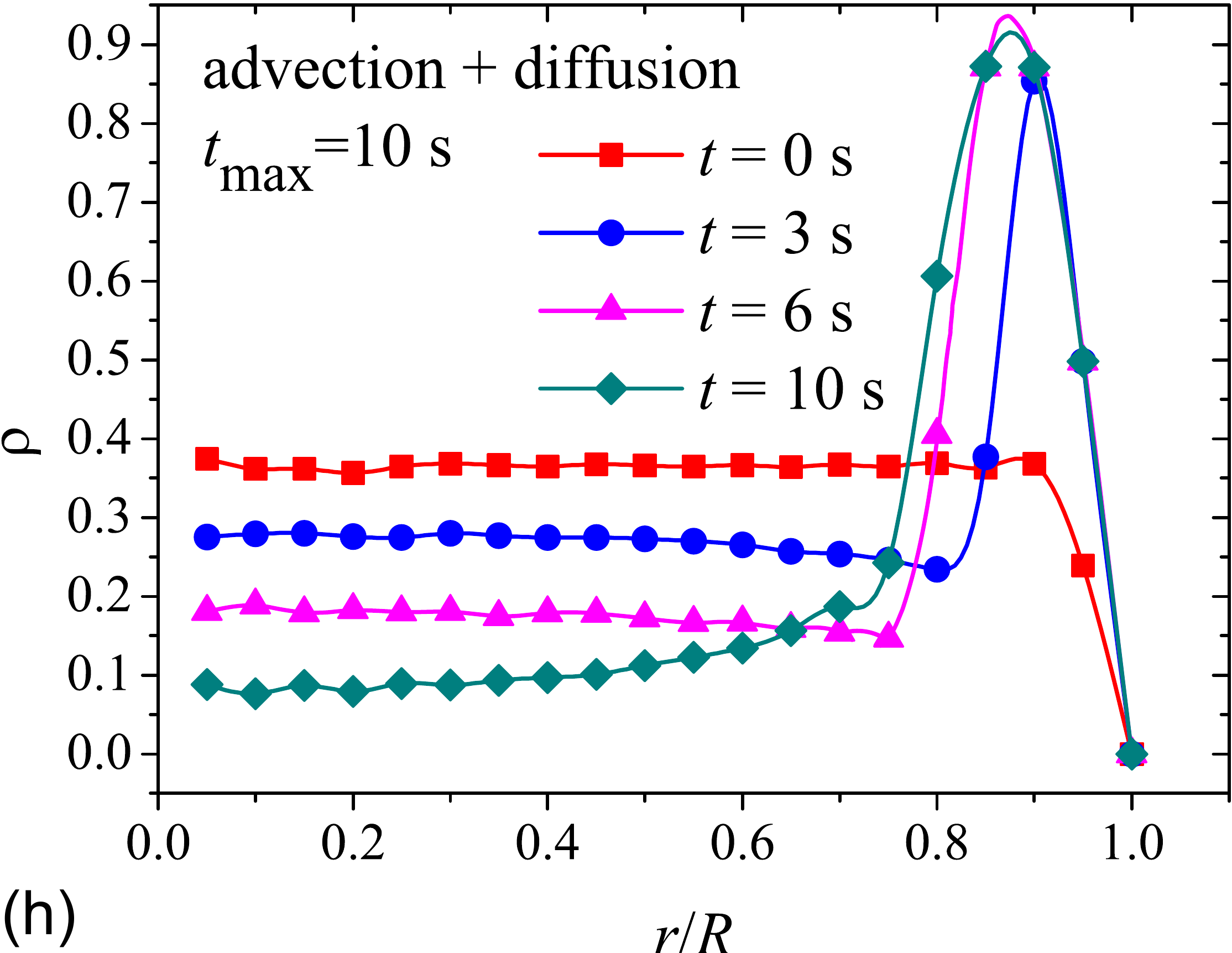}\\ \vspace{10pt}
     \includegraphics[height=4.25cm]{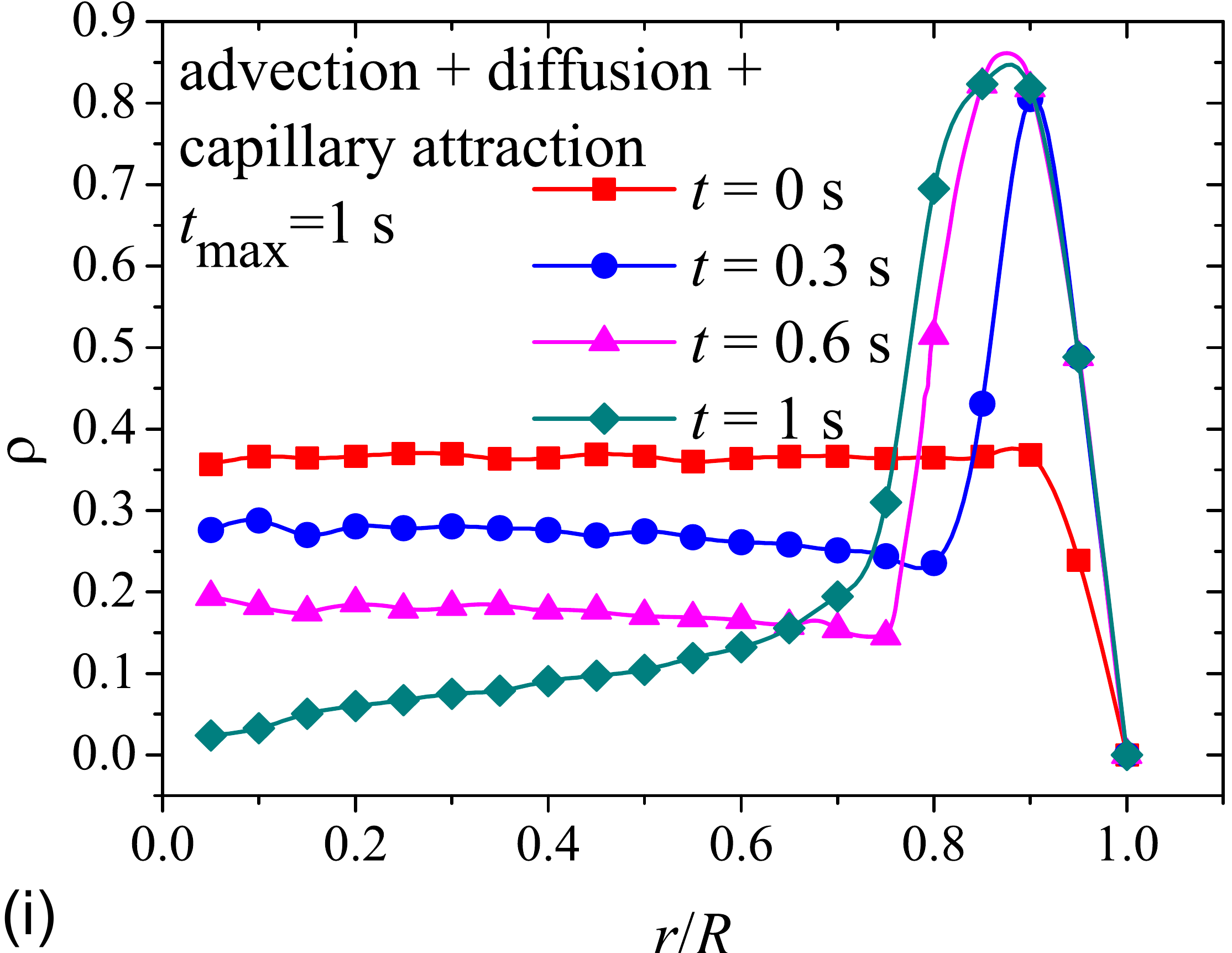}  \hspace{30pt}
     \includegraphics[height=4.25cm]{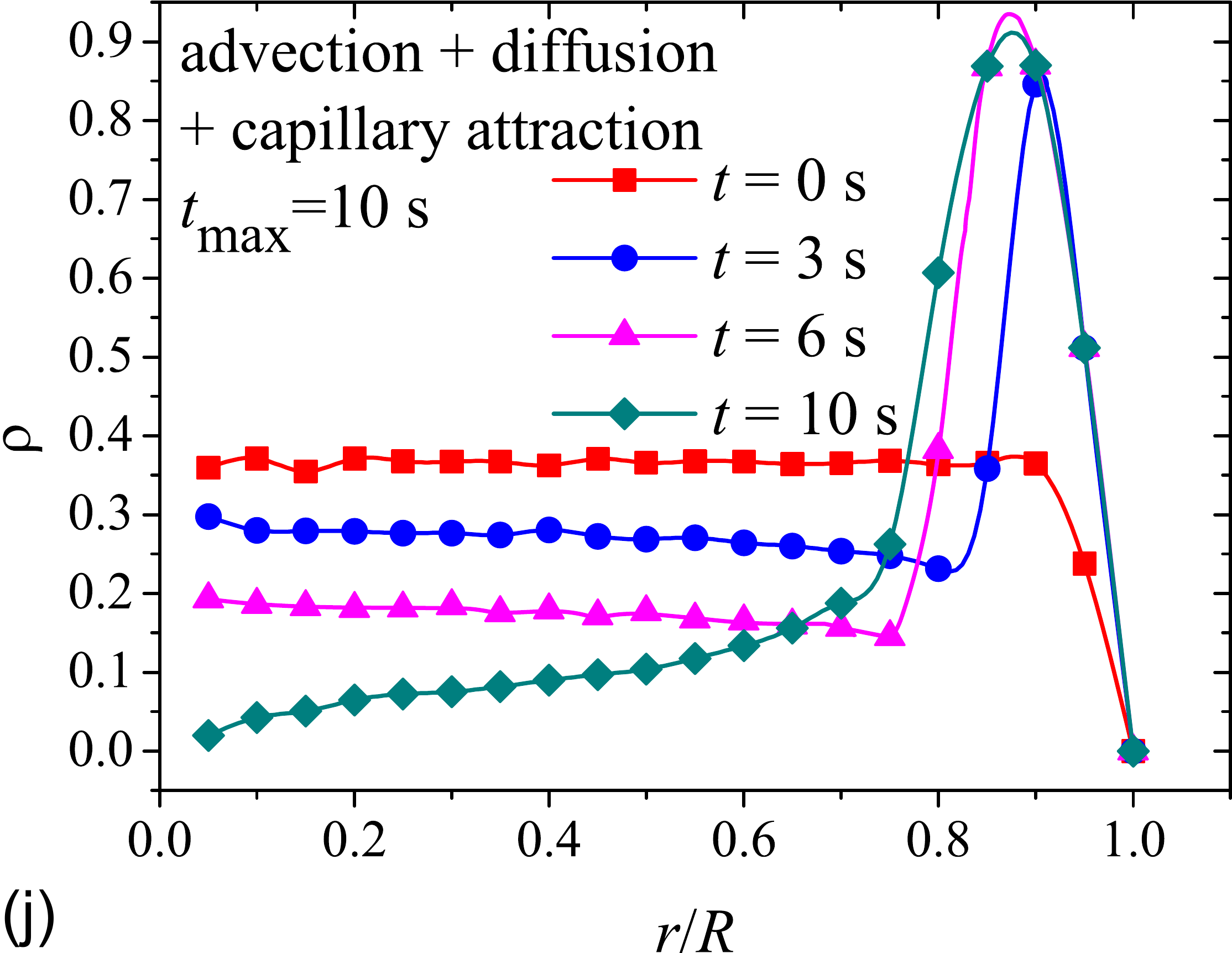}
     \caption{\label{fig:density} Density of the particles distribution for the obtained structures
     ($\rho$ is normalized to $N_p/R^2$).}
\end{figure*}

We chose the number of particles $N_p = 9000$ such that the width of
the formed annular deposition corresponded to the value from the
experiment~\cite{Park20063506}. The time $t_\mathrm{max}$ was
considered as a parameter of the simulations.
Figure~\ref{fig:structure} represents the obtained structures of
depositions at the drying time of $t_\mathrm{max} =$ 1~s (left) and
10~s (right). Figure~\ref{fig:density} represents the radial profile
of the particle density distribution $\rho$, which is one of the key
physical quantities studied. Here, $\rho$ is the number of particles
per unit area ($\rho$ is normalized to $N_p/R^2$). The profile is
demonstrated in two columns of panels for droplets with two
substantially different characteristic values of $t_\mathrm{max}$,
respectively. Five rows of panels in each of the two columns show
the radial profile calculated for five combinations of the main
effects considered, namely, calculated by taking into account solely
the advection [Figs.~\ref{fig:density}(a) and \ref{fig:density}(b)],
or the effects of diffusion and capillary attraction
[Figs.~\ref{fig:density}(c) and \ref{fig:density}(d)], or the
advection and capillary attraction [Figs.~\ref{fig:density}(e) and
\ref{fig:density}(f)], or the advection and diffusion
[Figs.~\ref{fig:density}(g) and \ref{fig:density}(h)], or the
advection, diffusion and capillary attraction jointly
[Fig.~\ref{fig:density}(i) and \ref{fig:density}(j)]. In parallel
with these ten panels, the panels of Fig.~\ref{fig:structure} show
the corresponding final spatial particle structures formed in each
case. Such a representation allows one to see the relative role of
each of the effects in forming the spatial particle structure on the
substrate at four characteristic times chosen, that correspond to
four curves in each of the panels in Fig.~\ref{fig:density}. The
calculation error does not exceed the size of the marker. Each
calculation was repeated ten times. The lines connecting the markers
are shown for convenience only and do not have a physical meaning.

Based on the simulation results, time $t_\mathrm{max}$ influences
the forming structure only in the cases when both diffusion and
advection are taken into account
[Figs.~\ref{fig:structure}(g)--\ref{fig:structure}(j)]. The fluid
flow carries the particles toward the periphery. A ring with
unordered particles is formed near the contact line over time
$t_\mathrm{max} =$ 1~s. Thus, if $t_\mathrm{max} < t_d$, the
particles do not have enough time to pack densely due to the
Brownian motion. We have calculated the packing fraction $p\approx
0.79$. The coffee rings formed during $t_\mathrm{max} =$ 10~s, are
characterized by almost hexagonal package of particles. In this
case,  the ratio of sediment area in the ring to ring area is
approximately 0.84. This ratio is about 0.91 in hexagonal packaging.
The difference is due to visible defects and local subdomains in the
obtained structure. In particular, the defects are induced by the
annular geometry of the region. Over $t_\mathrm{max} > t_d$, the
particles have enough time to densely pack due to diffusive mixing.
Thus, as a result of transfer of the particles by the advective flow
alone, the total density of the particles near the periphery is not
so high [$\rho \approx$ 0.56 in Figs.~\ref{fig:density}(a) and
\ref{fig:density}(b)], while under the additional influence of
diffusion $\rho \approx$ 0.81 for $t_\mathrm{max} =$ 1 s
[Figs.~\ref{fig:density}(g) and \ref{fig:density}(i)] and $\rho
\approx$ 0.86 for $t_\mathrm{max} =$ 10 s
[Figs.~\ref{fig:density}(h) and \ref{fig:density}(j)]. The influence
capillarity produces is that, instead of a uniform value of $\rho$
in the central part of the deposition [Fig.~\ref{fig:density}(g) and
\ref{fig:density}(h)], the final density distribution of the
particles grows linearly from the symmetry axis ($\tilde r = 0$)
toward the ring boundary ($\tilde r \approx 0.7$) in
Figs.~\ref{fig:density}(i) and \ref{fig:density}(j). We find that
$\rho(\tilde r = 1) = 0$, since the particles do not move beyond
boundary $R_f(0)$ toward $R$ (Fig.~\ref{fig:density}).

In the case of studying only advection or its combination with
capillarity, treelike structures of particles are observed within
the ring [Figs.~\ref{fig:structure}(a), \ref{fig:structure}(b),
\ref{fig:structure}(e), and \ref{fig:structure}(f)]. Such dendritic
shapes are not characteristic to diffusing particles. In the absence
of the Brownian motion, the particles do not mix within the ring. A
particle driven by the fluid flow stops moving when it encounters an
obstacle. Then, this particle can represent an obstacle for another
particle. This is the essence of the dendritic shape formation
mechanism.

If the system is only affected by diffusion and capillarity, the
depositions also have dendritic structures
[Fig.~\ref{fig:structure}(c) and \ref{fig:structure}(d)]. The
difference is that such structure occupies the entire surface area,
which was previously in contact with the liquid layer. Particle
density distribution is almost uniform ($\rho \approx 0.35$) in this
case [Fig.~\ref{fig:density}(c) and \ref{fig:density}(d)]. Besides,
the ``tree branches'' turn out sparse, since they are not formed due
to advection. The particles are uniformly distributed in the entire
volume and fluctuate chaotically at the beginning of the process. As
the fixing radius decreases, the particles occurring at the fixing
boundary ($\left| R_f - r \right| \leq r_p$), are drawn to each
other and stop their movement. They are in the state of rest for
several reasons. Firstly, the local droplet height is smaller than
the particle size, therefore, the particles lie on the substrate.
Secondly, the capillary forces press them to the substrate, because
a thin film wets these particles. At the beginning of the process,
the motion of the boundary $R_f$ is slow, and separate strips of
particles are formed. These strips are located near the periphery
perpendicular to the radial direction. There is a small dip in the
particle density distribution [Fig.~\ref{fig:density}(c) and
\ref{fig:density}(d)] right before the deposit, e.g., for $t = 0.6
t_\mathrm{max}$. The dip is located exactly at the fixing radius
area, where the local droplet height is comparable to the particle
size at each time, as can be seen in the Supplemental Material
video~\footnote{See Supplemental Material
at~\href{http://link.aps.org/supplemental/10.1103/PhysRevE.100.033304}{http://link.aps.org/\newline
supplemental/10.1103/PhysRevE.100.033304} for the simulation
videos}. The dip can be explained by the capillary attraction acting
in this small region, while the tree-like structure of particles is
still not finalized and the density is still increasing.

\begin{figure}[!htb]
 \includegraphics[width=0.95\linewidth]{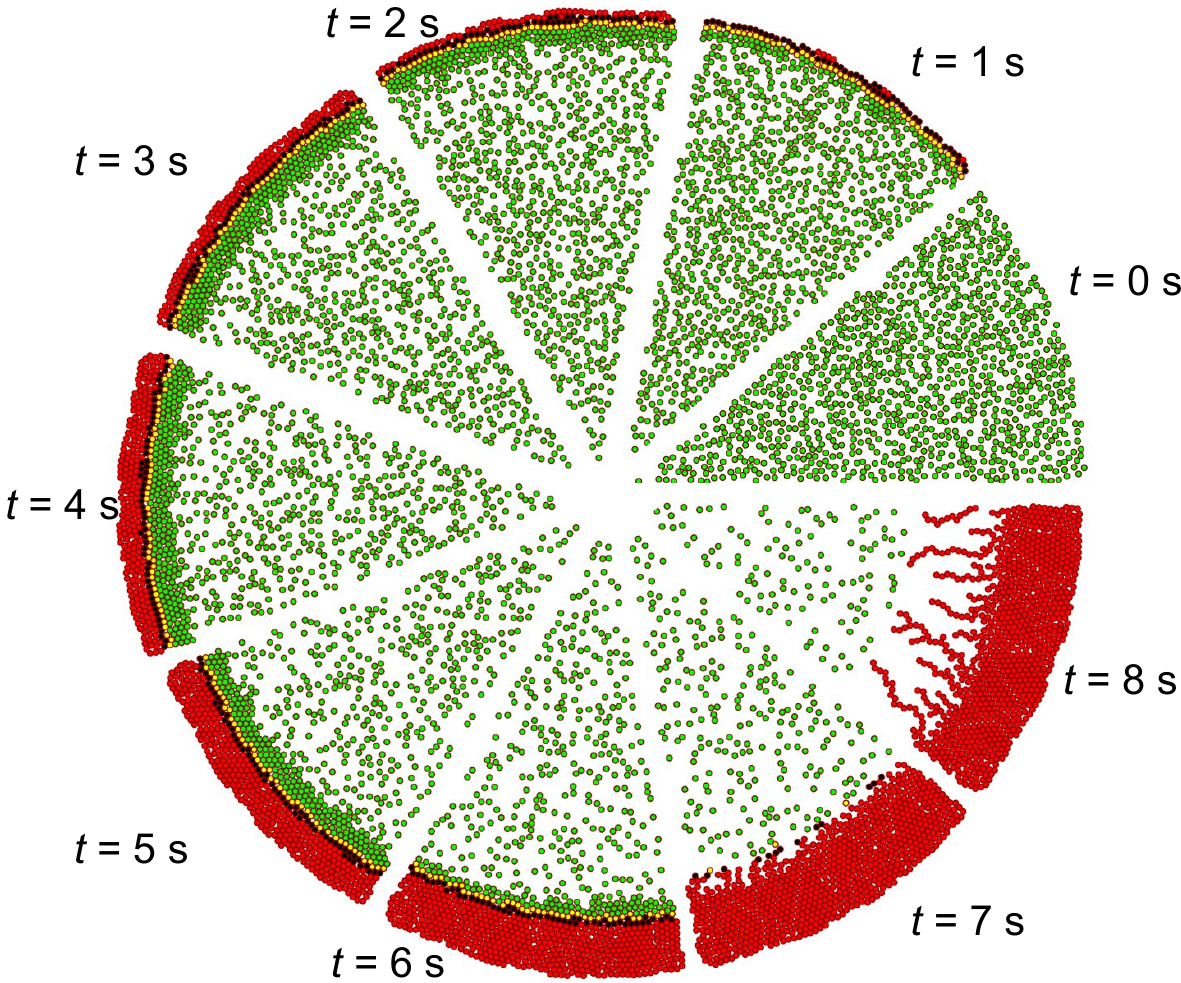}%
 \caption{\label{fig:pizza}
Time evolution of the particle distribution
for the case when advection,
 diffusion and capillary attraction are taken into account (color online).}
 \end{figure}

It should be noted that in cases when advection is taken into
account but capillary attraction is not, uniform distribution of the
particles is observed in the central region
[Fig.~\ref{fig:structure}(a), \ref{fig:structure}(b),
\ref{fig:structure}(g), and \ref{fig:structure}(h)]. These are the
particles not yet carried by the flow toward the annular deposition
at the droplet periphery. When capillarity is added to the model, we
see how these particles form chains [Fig.~\ref{fig:structure}(e),
\ref{fig:structure}(f), \ref{fig:structure}(i), and
\ref{fig:structure}(j)]. Capillarity does not affect the structure
inside the forming ring except drawing the particles, which were in
the proximity $R_f$, toward each other. Most of the time, the change
of $R_f(t)$ occurs slower than the growth of ring width $w(t)$. By
the end of the process, the situation becomes opposite, $R_f(t) <
R_f(0) - w(t)$ at $t > t_\mathrm{cri}$. Then the chains start
forming from the particles as they appear at fixing boundary $R_f$
(Fig.~\ref{fig:pizza}).

\begin{figure}[!htb]
 \includegraphics[width=0.75\linewidth]{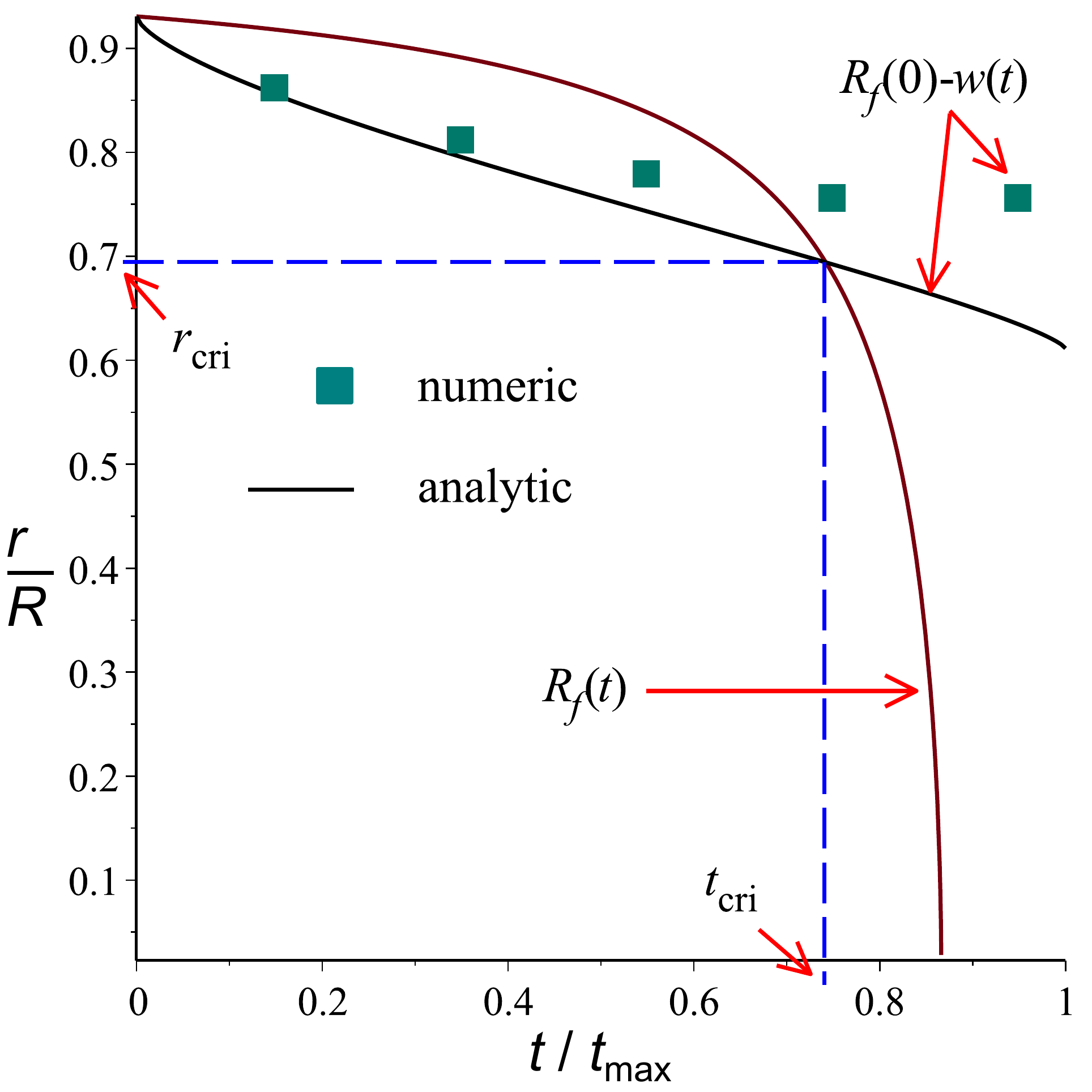}%
 \caption{\label{fig:Rf_Vs_w}
 Dynamics of the fixing radius and the inner front of the annular sediment.}
 \end{figure}

According to the numerical results, critical time $t_\mathrm{cri}
\approx (0.74 \pm 0.06) t_\mathrm{max}$. It is possible to obtain
the time $t_\mathrm{cri}$ from the following equality $R_f(
t_\mathrm{cri} )=R_f(0)-w( t_\mathrm{cri} )$. An analytical formula
for $w(t)$ was obtained in~\cite{Deegan2000475},
\begin{equation}
\label{eq:RingWidth} w= R \sqrt{\frac{\phi}{4p}} \left[ 1- \left( 1-
\frac{t}{t_\mathrm{max}} \right)^{3/4} \right]^{2/3},
\end{equation}
where $\phi$ is the volume fraction of colloidal particles. Let us
consider the critical time for a particular example, when three
effects affecting mass transfer are taken into account
($t_\mathrm{max}=1$~s). As shown in Fig.~\ref{fig:Rf_Vs_w},
the functions $R_f(t)$ and $R_f(0)-w(t)$ overlap
at $t_\mathrm{cri}\approx 0.74 t_\mathrm{max}$:
$R_f(t_{\mathrm{cri}})=
R_f(0)-w(t_{\mathrm{cri}})=r_\mathrm{cri}\approx 0.7 R$.
It is seen that the value
$t_\mathrm{cri}$ corresponding to the intersection of the $R_f(t)$ and
the numerically obtained $R_f(0)-w(t)$ is only slightly smaller than $0.74 t_\mathrm{max}$
(Fig.~\ref{fig:Rf_Vs_w}).

The result of modeling the dynamics of the
inner front of the annular sediment is quantitatively different from
the analytical prediction of~\eqref{eq:RingWidth} at the final
stage (Fig.~\ref{fig:Rf_Vs_w}). The value of the function $R_f(0)-w(t)$ at $t>t_\mathrm{cri}$
remains constant. This means that the width of the ring does not
increase further in time, as in the case of using the
formula~\eqref{eq:RingWidth}, because Eq.~\eqref{eq:RingWidth}
does not take into account the existence of the boundary $R_f$.

Branching in the polar direction is mostly suppressed in the region
where radial motion of the boundary $R_f$ is faster than the
characteristic velocity of advection and diffusion (60 $\mu$m/s),
while the particle density is low ($\rho < 0.2$) (see
Figs.~\ref{fig:structure}, \ref{fig:density}, and
\ref{fig:Rf_Vs_w}).

Considering jointly advection, diffusion, and capillarity for
$t_\mathrm{max} = 10$ s allowed obtaining a numerical result
[Fig.~\ref{fig:structure}(j)], which is qualitatively consistent
with the experimental data in Fig.~\ref{fig:ParkMoonExperiment}. The
structure of the deposition shows close-packed particles in the
annular part and snakelike chains of particles in the central
region. The difference is that these chains are short in the
experiment. The numerical results show long chains of particles. It
is most likely connected to the used approximation in the simulation
of the particles capillarity.

\section{Conclusions}

The formation of both close-packed annular deposition at the
periphery of the droplet and the snake-like chains of particles in
the central region has been described based on the suggested model.
 A major part of the colloidal
particles is transferred toward the periphery due to advection. The
fluid flow toward the contact line originates from the nonuniform
evaporation rate along the free surface of the droplet. This results
in the formation of the ring of colloid particles. The hexagonal
structure within the ring is found to be induced by the
diffusion-based ordering of the particles when the evaporation time
$t_\mathrm{max}$ exceeds the characteristic time $t_d$. Chains of
particles are formed at the final stage in the flattened droplet as
a result of the particles capillarity, when the fixing boundary
outreaches the inner boundary of the expanding ring, $R_f(t) <
R_f(0) - w(t)$.

The presented model jointly takes into account the effects of
advection, diffusion, and capillarity on the particle motion. We
consider only the case when the concentration of colloidal particles
and the droplet size allow a formation of a monolayer of particles.
 Our simplified model
and the corresponding numerical simulations allowed us to study some
of the effects discovered in the experiment~\cite{Park20063506}.
Since the results obtained are in a qualitative agreement with the
observations, we believe that the model in this case  accounts for
the major physical mechanisms of the deposition microstructure
formation.

\vspace{0.5cm} \acknowledgments

This work is supported by Grant No. 18-71-10061 from the Russian
Science Foundation. The authors would like to thank Yuri Yu.
Tarasevich, Irina V. Vodolazskaya, Renat K. Akhunzhanov, and Yury A.
Budkov for useful discussions and remarks.

\section{Appendix}\label{sec:appendix}
The pseudocode of the algorithm is given in the following
Algorithm~\ref{alg:particleDynamics}.
\begin{algorithm}[H]
\caption{Particle dynamics calculation} \label{alg:particleDynamics}
    \begin{algorithmic}[1]
      \State Set the parameters of the problem: $r_p$, $R$, $N_p$, and
      $t_\mathrm{max}$.
      \State Randomly generate a uniform distribution of particles and their coordinates $p[i].x$ and $p[i].y$, where $i \in [1, 2,\dots, N_p]$.
      \State By default, all particles are marked green.
      \For {$j \leftarrow 1, t_\mathrm{max}/\delta t$}
            \State Calculate $R_f$
            according to~\eqref{eq:fixing_radius}. To avoid dividing by zero
            (at $t \to t_\mathrm{max}$) $R_f \leftarrow r_p$,
            when $R_f < r_p$.
            \For {$i \leftarrow 1, N_p$}
                \State Override the color of the particle with regard to its position relatively to $R_f$.
                \State Skip the red particles.
                \If {(the particle is green or yellow)}
                    \State calculate new coordinates of the particles resulting from the diffusion.
                    \If {(no collision)}
                        \State move the particle.
                    \EndIf
                \EndIf
                \If {(the particle is green)}
                    \State calculate new coordinates of the particles resulting from the advection.
                    \If {(no collision)}
                        \State move the particle.
                    \EndIf
                \EndIf
                \If {(the particle is black and there is a neighboring particle that is either black or red in its surroundings $R_n$)}
                    \State calculate new coordinates of the particles resulting from the capillarity.
                    \If {(no collision)}
                        \State place the current particle closely to the neighboring one.
                    \EndIf
                \EndIf
            \EndFor
            \State Write the coordinates of the particles and their colors for the current time step to a file.
      \EndFor
    \end{algorithmic}
\end{algorithm}

\bibliography{biblio}

\end{document}